\documentclass[aps,prd,amsfonts,amssymb,amsmath,twocolumn]{revtex4}

\usepackage{graphicx}
\usepackage{dcolumn}
\usepackage{bm}

\begin{document}

\title{Some general properties of the renormalized stress-energy
tensor for static quantum states on $(n+1)$-dimensional spherically symmetric black holes}

\author{Dean Morgan}
\author{Stuart Thom}
\author{Elizabeth Winstanley} \email{E.Winstanley@sheffield.ac.uk}
\author{Phil M. Young}
\affiliation{Department of Applied Mathematics, The University of
Sheffield,
Hicks Building, Hounsfield Road, Sheffield, S3 7RH, United Kingdom.}

\date{\today}


\begin{abstract}
We study the renormalized stress-energy tensor (RSET) for static quantum states on
$(n+1)$-dimensional, static, spherically symmetric black holes.
By solving the conservation equations, we are able to write the stress-energy
tensor in terms of a single unknown function of the radial co-ordinate, plus
two arbitrary constants.
Conditions for the stress-energy tensor to be regular at  event horizons
(including the extremal and ``ultra-extremal'' cases) are
then derived using generalized Kruskal-like co-ordinates.
These results should be useful for future calculations of the RSET for
static quantum states on spherically symmetric black hole geometries in any
number of space-time dimensions.
\end{abstract}


\maketitle

\section{Introduction}
\label{sec:intro}

The renormalized stress-energy tensor (RSET) $\langle T_{\mu \nu } \rangle _{\rm {ren}}$
is an object of fundamental
importance in quantum field theory in curved space-time, since it governs, via
the semi-classical Einstein equations
\begin{equation}
G_{\mu \nu } = 8\pi G \langle T_{\mu \nu } \rangle _{\rm {ren}},
\label{eq:scEE}
\end{equation}
the back-reaction of the quantum
field on the space-time geometry.
However, the renormalization process means that
detailed calculations of the RSET
are notoriously difficult, even in spherically symmetric space-times
(see, for example, \cite{ash} for some four-dimensional calculations),
although calculations in three space-time dimensions are more tractable \cite{3D}.
Various analytic approximations \cite{approx} have been developed for the RSET
in various cases, but it is still useful to obtain as much information as possible about the
RSET from basic physical principles without
a full calculation.

Christensen and Fulling \cite{cf} pioneered this approach.
From equation (\ref{eq:scEE}), since the Einstein tensor $G_{\mu \nu }$ is
divergence-free, the same must be true of
$\langle T_{\mu \nu } \rangle _{\rm {ren}}$.
Christensen and Fulling \cite{cf} therefore studied solutions of the conservation
equations
\begin{equation}
\nabla ^{\mu } \langle T_{\mu \nu } \rangle _{\rm {ren}}=0
\label{eq:cons}
\end{equation}
on the Schwarzschild black hole geometry, in both two and four space-time dimensions.
In four dimensions, the conservation equations, together with elementary symmetry principles,
can be solved to give the renormalized stress-energy tensor in terms of one unknown function
and two unknown constants.
The constants are constrained by the particular choice of vacuum state
(Hartle-Hawking \cite{hh}, Unruh \cite{u} or Boulware \cite{b}), through the regularity
properties of the RSET on the event horizon.
In two dimensions, the stress-energy tensor is given completely in terms of the trace anomaly
{\footnote{The conformal anomaly has also been used in four dimensions to study the back-reaction,
see, for example, \cite{nojiri}.}}.
The analysis of \cite{cf} has proved to be powerful for reducing the calculation of
the RSET to a single component (typically taken to be $T_{\theta }^{\theta }$), but also because
it revealed certain properties of the RSET (such as its singularity structure on the horizons)
without requiring a full computation.

There is now great interest in a wide variety of static black hole
space-times, and it is our purpose in this paper to extend the analysis of \cite{cf} to
more general, $(n+1)$-dimensional, static black holes.
Our analysis is independent of the asymptotic structure of the geometry at infinity, and
therefore includes ``topological'' black holes which can exist in asymptotically
anti-de Sitter space \cite{topological} as well as general, spherically
symmetric black holes.
However, in this paper, for ease of phraseology
we will tend to use the phrase ``spherically symmetric'' even when our results
include these topological black holes.
A key question in any study of the RSET is whether or not it is regular across a horizon.
We examine this for a general horizon (whether non-extremal or extremal), using
generalized Kruskal-like co-ordinates \cite{liberati} which can describe both extremal and non-extremal
horizons.

Our analysis in this paper is restricted to static quantum states.
This is sufficiently general to cover the Hartle-Hawking \cite{hh} and Boulware \cite{b}
vacua for all static black hole geometries.
These two vacua (particularly the Hartle-Hawking state)
are the ones most frequently studied in full calculations of the RSET \cite{ash},
as they are time-reversal symmetric and therefore possess the greatest symmetries, which
makes calculations easier.
Our results will certainly be of use for future calculations of the RSET in these two states.
However, the Unruh \cite{u} vacuum may not be covered by our analysis in general.
For static, spherically symmetric, asymptotically flat black holes with a non-extremal event horizon,
it is likely that the Unruh vacuum will be a static state and therefore covered by our approach.
However, it is known that for black holes with extremal horizons \cite{extremal}, or with
both event and cosmological horizons \cite{sds,choudhury}, the equivalent of the Unruh vacuum is not a static state.
For non-static states, the conservation equations are considerably more complex and correspondingly less
information is accessible from their solution, apart from in two space-time dimensions
(a similar situation arises if one studies the solutions
of the conservation equations on a Kerr black hole \cite{aco}).
We therefore do not consider this situation further.

The outline of this paper is as follows.
In section \ref{sec:geometry} we describe our static, black hole metric, and
our generalized Kruskal-like co-ordinates (following \cite{liberati}) which are regular across any horizon.
The conservation equations (\ref{eq:cons}) are solved on this background metric in
section \ref{sec:conseqns}.
This gives the RSET in terms of a single unknown function of the radial co-ordinate $r$ and
two arbitrary constants.
We then turn, in section \ref{sec:RSEThor},
to the regularity of the RSET across an arbitrary horizon, and derive conditions for the RSET to be regular.
These conditions are particularly stringent when the horizon is extremal.
In section \ref{sec:eventandcosmo} we focus on the special case of a space-time with
distinct event and cosmological horizons, and derive a strong integral constraint
which must be satisfied if the RSET is to be regular across both the event and cosmological
horizons.
Our focus in this paper is the study of $(n+1)$-dimensional black holes, however in
section \ref{sec:2d}, we restrict attention to the case of two space-time dimensions,
where the unknown function of $r$ disappears.
For distinct event and cosmological horizons, we show that the RSET can be regular
across all horizons only if they have the same temperature, and we
are also able to show that no static state can have an RSET which is regular across an extremal horizon.
Finally, our conclusions are presented in section \ref{sec:conc}.
Henceforth, the metric has signature $(-,+,+,+)$ and we use units in which $c=G=\hbar =1$.

\section{General $(n+1)$-dimensional spherically symmetric black holes}
\label{sec:geometry}

\subsection{Metric ansatz}
\label{sec:metric}

We consider an $(n+1)$-dimensional, static,
black hole metric in the following form, after a suitable choice of gauge:
\begin{equation}
ds^{2}= -f(r) dt^{2} + f^{-1}(r) dr^{2} + R^{2}(r) \, d\Omega _{n-1}^{k},
\label{eq:metric}
\end{equation}
where the metric functions $f$ and $R$ depend on the radial co-ordinate $r$ only.
The $(n-1)$-dimensional metric $ d\Omega _{n-1}^{k}$ (where we have
used the notation of \cite{cardoso}) is as follows, for $k=1$, $k=0$ and $k=-1$
respectively:
\begin{eqnarray}
d\Omega^{1}_{n-1} & = &
dx_{2}^{2}+\sin^{2}x_{2}\,dx_{3}^2+....
+\prod_{i=2}^{n-1}\sin^{2}x_{i}\, dx_{n}^2;
\nonumber \\
d\Omega^{0}_{n-1} & = &
dx_{2}^{2}+dx_{3}^2+dx_{4}^2+....+dx_{n}^2 ;
\nonumber \\
d\Omega^{-1}_{n-1} & = &
dx_{2}^{2}+\sinh^{2}x_{2}\,dx_{3}^2+....
\nonumber \\ & &
+\prod_{i=2}^{n-1}\sinh^{2}x_{i}\, dx_{n}^2 .
\end{eqnarray}
We can only have $k\neq 1$ for asymptotically adS (anti-de Sitter) black holes.
The particular value of $k$ describes the topology of the
event horizon, which is spherical for $k=1$; planar,
cylindrical or toroidal (with genus $\geq 1$) for $k=0$;
and hyperbolic or toroidal (with genus $\geq 1$) for
$k=-1$.
Although not strictly speaking correct, in the remainder of the paper we shall use the
phrase ``spherically symmetric'' to include these topological black holes.

For most of the exact
solutions in which one might be interested (for example, the higher-dimensional
Reissner-Nordstr\"om(-anti)-de Sitter black holes), it will be the case that $R(r)\equiv r$,
which simplifies the analysis.
However, our metric (\ref{eq:metric}) is sufficiently general to cover many black hole
solutions in dilaton gravity, hairy black holes, and even the Nariai \cite{nariai} metric
(in which case $R\equiv \Lambda >0$).
The general $(n+1)$-dimensional, spherically symmetric metric is also commonly
written in the form
\begin{equation}
ds^{2} = - N(R) S^{2}(R) \, dt^{2}+  N(R)^{-1} dR^{2} + R^{2} d\Omega _{n-1}^{k}.
\end{equation}
This metric can be transformed into the form (\ref{eq:metric}) by the change of co-ordinates
\begin{equation}
\frac {dr}{dR} = S(R).
\end{equation}

We will be particularly interested in the regularity of the RSET at
a horizon of the black hole geometry.
Assuming that $R(r)$ has no zeros, the horizon structure
of the black hole is determined by the metric function $f(r)$.
We will assume that $f(r)$ has at least one zero, namely a regular black hole event
horizon at $r=r_{+}$, with $f'(r_{+})>0$.
In the presence of a positive cosmological constant, there will also be a
cosmological event horizon at $r=r_{++}$, where $f'(r_{++})<0$.
Our analysis also allows the possibility of an inner, Cauchy, horizon at $r=r_{-}$,
with $f'(r_{-})<0$.
It is possible for two or more of these horizons to coincide.
Therefore, there are several different types of horizon which we need to consider:
\begin{enumerate}
\item
Regular, non-extremal horizons, of event, cosmological or inner variety
(at $r_{+}$, $r_{++}$ or $r_{-}$ respectively);
\item
An extremal horizon, formed by the coincidence of an event and inner horizon
(the so-called ``cold'' black hole \cite{romans});
\item
An extremal horizon, formed by the coincidence of an event and cosmological horizon
(a ``Nariai'' black hole \cite{cardoso});
\item
An ``ultra-extremal'' black hole horizon, formed by the coincidence of all three types of horizon
(the ``ultra-cold'' black hole \cite{romans}).
\end{enumerate}
All these possibilities can be illustrated by solutions of Einstein-Maxwell theory \cite{cardoso}.
In these specific examples, $R(r)=r$, and the metric function $f(r)$ is given by:
\begin{equation}
f(r)=k-\frac {M}{r^{n-3}}+
\frac {q^{2}}{r^{2(n-3)}}
-\frac {\Lambda r^{2}}{3},
\label{eq:fRNdS}
\end{equation}
where $M$, $q$ are related, respectively, to the mass and charge of the black hole \cite{eugen}.
Penrose diagrams for all these possible cases can be found in \cite{cardoso}.

\subsection{Kruskal-like co-ordinates}
\label{sec:kruskal}

In order to analyze the regularity of the RSET at the horizons of the spacetime,
we require Kruskal-like co-ordinates which are regular across each horizon.

We begin by defining the usual
``tortoise'' co-ordinate $r^{*}$ by the equation:
\begin{equation}
\frac {dr^{*}}{dr} = \frac {1}{f(r)},
\label{eq:rstar}
\end{equation}
so that an event horizon $r=r_{+}$ corresponds to $r_{*}\rightarrow -\infty $.
If there is a cosmological horizon $r=r_{++}$, then it will be the case that
$r_{*}\rightarrow \infty $ there, as is also true as $r\rightarrow \infty $
for asymptotically flat spacetimes.
However, if the geometry is asymptotically adS, then $r_{*}$ tends to a finite constant
(which we may as well take to be zero) at infinity.

Starting with the usual null co-ordinates
\begin{equation}
v=t+r^{*}, \qquad w = t-r^{*},
\label{eq:nullcoords}
\end{equation}
for all types of non-extremal horizon, we define the standard Kruskal co-ordinates, $V$, $W$,
in a region in which $f(r)>0$, by
\begin{equation}
V = e^{\kappa v}, \qquad
W = -e^{-\kappa w} ,
\label{eq:kruskal}
\end{equation}
where the surface gravity $\kappa $ is given by
\begin{equation}
\kappa = \frac {1}{2} f'(r_{0}) ,
\label{eq:kappa}
\end{equation}
with $r_{0}$ being the location of the particular horizon under consideration.
However, at an extremal horizon, the surface gravity $\kappa $ (\ref{eq:kappa}) vanishes
and so the standard Kruskal co-ordinates (\ref{eq:kruskal}) are constant.
In this situation one can use Eddington-Finkelstein co-ordinates in patches
across the future and past horizons separately,
but we shall instead follow the method of \cite{liberati} to define new
Kruskal-like co-ordinates in this case.
In \cite{liberati}, new co-ordinates were defined in the case of coincident event and
inner horizons, and we here extend their method to the general situation.
In addition, these new co-ordinates can be defined equally well for non-extremal horizons, which
will allow us, in section \ref{sec:RSEThor}, to deal simultaneously with
the analysis of the behaviour of the RSET near all types of horizon.

At an extremal horizon, the metric function $f(r)$ will have either a double or a triple
zero at $r=r_{0}$ (the triple zero occurring in the ``ultra-extremal'' case), and
we therefore write $f(r)$ as
\begin{eqnarray}
f(r) & = & (r-r_{0}) \, g_{1}(r);
\nonumber \\
f(r) & = & (r-r_{0})^{2} g_{2}(r) ;
\nonumber \\
f(r) & = & (r-r_{0})^{3} g_{3}(r) ;
\end{eqnarray}
for a non-extremal, extremal and ultra-extremal horizon respectively,
where $g_{1}(r)$, $g_{2}(r)$, $g_{3}(r)$ are non-zero at $r=r_{0}$.
Integrating (\ref{eq:rstar}) gives, near the horizon,
\begin{eqnarray}
r_{*} & = &
a_{1} \log (r-r_{0}) + O(1);
\nonumber \\
r_{*} & = &
b_{1} (r-r_{0})^{-1} + b_{2} \log (r-r_{0}) + O(1);
\nonumber \\
r_{*} & = &
c_{1} (r-r_{0})^{-2} + c_{2} (r-r_{0}) ^{-1}
\nonumber \\ & &
+ c_{3} \log (r-r_{0}) + O(1),
\label{eq:rstarext}
\end{eqnarray}
for non-extremal,  extremal and ultra-extremal horizons, respectively, where
the $a$s, $b$s and $c$s are constants given in terms of the $g$s and their derivatives
at $r=r_{0}$.
The ones we need later are:
\begin{eqnarray}
a_{1} & = & g_{1}(r_{0})^{-1},
\nonumber \\
b_{1} & =  & -g_{2}(r_{0})^{-1},
\nonumber \\
c_{1} & = & -\frac {1}{2} g_{3}(r_{0})^{-1}.
\label{eq:abc}
\end{eqnarray}
Near the horizon, we have $r_{*} \rightarrow \pm \infty $, with the sign depending on the sign of $a_{1}$,
$b_{1}$ and $c_{1}$.

Following \cite{liberati}, we now define a function $\psi (\xi )$ as
one half of that
part of $r_{*}$ (\ref{eq:rstarext}) which is singular as $r\rightarrow r_{0}$, that
is, for non-extremal, extremal and ultra-extremal black holes respectively:
\begin{eqnarray}
\psi (\xi ) & = &  \frac {1}{2} a_{1} \log \xi ;
\nonumber \\
\psi (\xi ) & = & \frac {1}{2} \left(
b_{1} \xi ^{-1} + b_{2} \log \xi  \right) ;
\nonumber \\
\psi (\xi ) & = & \frac {1}{2} \left(
c_{1} \xi ^{-2} + c_{2} \xi  ^{-1}
+ c_{3} \log \xi   \right) .
\label{eq:psiform}
\end{eqnarray}
We then define new Kruskal-like co-ordinates ${\cal {V}}$, ${\cal {W}}$ implicitly by \cite{liberati}:
\begin{equation}
v = \psi ({\cal {V}}), \qquad
w = -\psi ({\cal {W}}) ,
\label{eq:genKrusk}
\end{equation}
where $v$ and $w$ are the null co-ordinates given in equation (\ref{eq:nullcoords}).
Note that the definition (\ref{eq:genKrusk}) reduces to the standard Kruskal co-ordinates (\ref{eq:kruskal})
for a non-extremal horizon.
In terms of these new co-ordinates, the metric (\ref{eq:metric}) takes the form
\begin{equation}
ds^{2} = -f(r) \psi ' ({\cal {V}}) \psi ' (-{\cal {W}}) \, d{\cal {V}} \,
d{\cal {W}} + R(r)^{2}  d\Omega _{n-1}^{k}.
\end{equation}

In order to show that ${\cal {V}}$ and ${\cal {W}}$ are in fact good co-ordinates across
the horizons, we need to consider future and past horizons separately.
The same argument works for both, so suppose we are considering a horizon ${\cal {H}}$
on which ${\cal {V}}$ is finite and non-zero, so that $\psi '({\cal {V}})$ is
also finite and non-zero there.
For many black holes, this will correspond to a future event horizon, where $t\rightarrow \infty $
and $r_{*} \rightarrow - \infty $, although this will depend on the signs of the constants in (\ref{eq:abc}).
The precise causal structure of the event horizon does not affect our construction.
Then $t+r_{*}$ is finite and non-zero on ${\cal {H}}$, giving, from the definition
of $\psi $,
\begin{equation}
t-r_{*} = -2r_{*}+O(1) = - \psi (r-r_{0}) + O(1).
\end{equation}
Therefore we have
\begin{eqnarray}
{\cal {W}} & = & - \psi ^{-1} (-t+r_{*})
= -\psi ^{-1} \left[ \psi (r-r_{0}) + O(1) \right]
\nonumber \\
& = & -(r-r_{0}) + O(1).
\label{eq:calWhor}
\end{eqnarray}
From the definition of $\psi $ in the non-extremal, extremal and ultra-extremal cases, respectively,
\begin{eqnarray}
\psi ' ( -{\cal {W}}) & = &
\frac {1}{2} g_{1} (r_{0}) (r-r_{0}) ^{-1} + O(1);
\nonumber \\
\psi ' (-{\cal {W}})  & = &
\frac {1}{2} g_{2}(r_{0})^{-1} (r-r_{0}) ^{-2} + O(r-r_{0})^{-1};
\nonumber \\
\psi ' (-{\cal {W}}) & = &
\frac {1}{2} g_{3}(r_{0})^{-1} (r-r_{0}) ^{-3} + O(r-r_{0})^{-2}.
\label{eq:psi'W}
\end{eqnarray}
In all three cases, then, $f(r)\psi '(-{\cal {W}})$ is finite and non-zero as
$r\rightarrow r_{0}$, so that ${\cal {V}}$ and ${\cal {W}}$ are suitable co-ordinates,
because $\psi '({\cal {V}})$ is also finite and non-zero.
Substituting ${\cal {V}}$ for ${\cal {W}}$ in the above argument shows that
these are also suitable regular co-ordinates across a horizon where ${\cal {W}}$ is finite and non-zero.

For spacetimes with many distinct horizons, patches of different
Kruskal co-ordinates may be required (see, for example, \cite{bazanski}).
In the particular case of a black hole with a regular event horizon at $r=r_{+}$ and
cosmological horizon at $r=r_{++}$, one patch of Kruskal co-ordinates, $V_{+}$, $W_{+}$,
can be used to cover the region extending from inside the event horizon up to the
cosmological horizon, but they will not be regular across the cosmological horizon.
We therefore need a second set of Kruskal co-ordinates, $V_{++}$, $W_{++}$, which are regular
across the cosmological horizon all the way down to the event horizon, but not across the
event horizon.

In \cite{choudhury,lake},
a co-ordinate system is given which is regular across {\em {both}} the
event and cosmological horizon in Schwarzschild-de Sitter space-time, and which
therefore removes the need to use two sets of Kruskal co-ordinates.
We have not used this here because we wish to work as generally as possible, and,
although the method of \cite{choudhury,lake} could be used to find globally regular
co-ordinates in a more general case, the expressions involved are likely to be
algebraically highly complex and make analysis difficult.

\section{Solution of the conservation equations}
\label{sec:conseqns}

Since we are working on a static, spherically symmetric black hole
spacetime, we will assume that the RSET (henceforth denoted simply
by $T_{\mu \nu }$) is also static and spherically symmetric.
The form of the RSET is then:
\begin{equation}
\label{eq:bsttr}
T^{\mu }_{\nu } =
\left(
\begin{array}{ccccc}
A & -Pf^{-1} &  &  &  \\
Pf & T-A-(n-1)Q & & &
\\  & & Q & &
\\ & & & \ddots &
\\  & & & & Q
\end{array}
\right) ;
\end{equation}
with all other entries vanishing,
where $A, P, Q$ and $T$ are functions of $r$ only.
The RSET will be symmetric under time-reversal symmetry if and only if $P\equiv 0$.

The usual trace anomaly is given by $T_{\alpha }^{\alpha }= T$.
For general $(n+1)$-dimensional spacetimes, the trace anomaly is
zero if $n$ is even, and, if $n$ is odd, is given in terms of the appropriate
DeWitt-Schwinger coefficient.
For example, for a massless, conformally coupled scalar field,
the result is \cite{christensen}:
\begin{equation}
T = \frac {1}{\left( 4\pi \right) ^{(n+1)/2}} {\mbox {Tr }} a_{(n+1)/2} .
\end{equation}
Various DeWitt-Schwinger coefficients have been calculated, giving the
trace anomaly in various spacetime dimensions \cite{dewitt}.
The trace anomaly is always a geometric scalar, and so is finite everywhere
apart from at a curvature singularity.
It is independent of the state of the quantum field under consideration, but
does depend on the spin of the quantum field.

The conservation equations (\ref{eq:cons}) arising from the $x_{i}$ co-ordinates
are trivial, and the $t$ and $r$ equations give, respectively,
\begin{eqnarray}
0 & = &
\frac {d}{dr} \left( f P R^{(n-1)} \right) ;
\nonumber \\
0 & = &
\frac{1}{R^{n-1}} \frac{d}{dr} \left( R^{n-1}fA \right)
+ \frac{(n-1)}{2QR^{2n}}
\frac{d}{dr} \left( Q^{2}fR^{2n}\right)
\nonumber \\ & &
-\frac{1}{2TR^{2(n-1)}}
\frac{d}{dr} \left(T^{2}fR^{2(n-1)}\right) ;
\end{eqnarray}
which can be readily integrated to give
\begin{eqnarray}
P & = &
\frac{X}{fR^{(n-1)}};
\nonumber \\
A & = &
-(n-1)Q+T +\frac{Z}{fR^{(n-1)}} +J(r_{a},r);
\label{eq:PA}
\end{eqnarray}
where $X$, $Z$ are integration constants and we define
\begin{eqnarray}
J(x,y) & = &
\frac{1}{2fR^{(n-1)}}
\int_{x}^{y}
\left[(n-1)Q-T \right]
f'R^{(n-1)} dr
\nonumber \\ & &
-\frac{(n-1)}{fR^{(n-1)}}
\int_{x}^{y}
QfR^{(n-2)} R' dr ,
\label{eq:Jdef}
\end{eqnarray}
with $r_{a}$ being any fixed value of $r$ (to be chosen shortly).
The formulae (\ref{eq:PA}) reduce to those in \cite{cf} when $n=1$ or $n=3$, and the
metric (\ref{eq:metric}) is Schwarzschild.
Here we find, like \cite{cf}, that the complete stress-energy tensor is given in terms of
two unknown constants $X$, $Z$ and one unknown function of $r$, which we can take to be $Q$.
If there are only two spacetime dimensions, then the RSET is determined solely by the trace
anomaly and the constants $X$ and $Z$ \cite{cf}.

\section{Behaviour of the RSET near a horizon}
\label{sec:RSEThor}

We now address the key question of the behaviour of the RSET close to horizons.
For all types of horizon at $r=r_{0}$, we employ the modified Kruskal-like co-ordinates
${\cal {V}}$, ${\cal {W}}$ constructed in section \ref{sec:kruskal}.
The relevant RSET components in these co-ordinates are:
\begin{eqnarray}
T_{ {\cal {V}} {\cal {V}} } & = &
\frac {1}{2} f \left[ \psi ' ({\cal {V}}) \right] ^{2}
\left\{ {\cal {F}}
- \frac {1}{fR^{(n-1)}} \left[ Z-X \right] \right\} ;
\nonumber \\
T_{ {\cal {W}} {\cal {W}} } & = &
\frac {1}{2} f \left[ \psi ' (-{\cal {W}}) \right] ^{2}
\left\{
{\cal {F}}
- \frac {1}{fR^{(n-1)}} \left[ Z+X \right] \right\} ;
\nonumber \\
T_{ {\cal {V}} {\cal {W}} } & = &
\frac {1}{4} f \psi ' ({\cal {V}}) \psi ' (-{\cal {W}})
\left\{ (n-1) Q - T \right\} ;
\label{eq:Tcalhor}
\end{eqnarray}
where
\begin{equation}
{\cal {F}}=\frac {1}{2} (n-1) Q - \frac {1}{2} T - J(r_{0},r)
\label{eq:calF}
\end{equation}
and we have chosen the lower limit in the definition of $J(r_{a},r)$ (\ref{eq:Jdef})
to be the location of the horizon, namely $r_{a}=r_{0}$.

Consider firstly a horizon on which ${\cal {V}}$ is finite
and non-zero, and ${\cal {W}}$ vanishes from (\ref{eq:calWhor}).
From section \ref{sec:kruskal}, on this horizon $\psi '({\cal {V}})$ is
also finite and non-zero, and, from (\ref{eq:psi'W}), we have
\begin{equation}
\psi ' (-{\cal {W}}) f = O(1)
\end{equation}
as $r\rightarrow r_{0}$.
From (\ref{eq:Tcalhor}), it is immediately clear that
$T_{ {\cal {V}} {\cal {W}} } = T_{ {\cal {W}} {\cal {V}} }$ is regular on this
horizon provided that $Q$ is; which we shall assume to be the case.
For the components $T_{ {\cal {V}} {\cal {V}} }$ and $T_{ {\cal {W}} {\cal {W}} }$
we need to analyze the behaviour of ${\cal {F}}$ (\ref{eq:calF})
as $r\rightarrow r_{0}$.
We assume that $f(r)$ has the form
\begin{equation}
f(r)= f_{p} \left( r-r_{0} \right) ^{p} + f_{p+1} \left( r-r_{0} \right) ^{p+1} + \ldots ;
\end{equation}
for $r\sim r_{0}$, where we are interested particularly in the cases
$p=1$ (non-extremal black hole),
$p=2$ (``cold'' \cite{romans}
black hole or ``Nariai'' \cite{cardoso} black hole) and
$p=3$ (``ultra-cold'' \cite{romans} black hole).
Performing a Taylor series expansion of all the quantities,
we find
\begin{equation}
{\cal {F}} (r) = K_{1} (r-r_{0}) + K_{2} (r-r_{0})^{2} + O(r-r_{0})^{3} ;
\end{equation}
where
\begin{eqnarray}
K_{1} & = &
\frac {1}{2(p+1)}
\left[
(n-1) {\tilde {\cal {G}}}(r_{0}) \frac {R'(r_{0})}{R(r_{0})}
+ {\cal {G}}'(r_{0})
\right] ;
\nonumber \\
K_{2} & = & \frac {1}{2(p+1)(p+2)} \left\{
2 (n-1) Q'(r_{0}) \frac {R'(r_{0})}{R(r_{0})}
\right. \nonumber \\ & &
+(p+1) {\cal {G}}''(r_{0})
-{\cal {G}}'(r_{0}) \frac {f_{p+1}}{f_{p}}
\nonumber \\ & &
+ p(n-1) {\tilde {\cal {G}}}'(r_{0}) \frac {R'(r_{0})}{R(r_{0})}
\nonumber \\ & &
+  (n-1) {\tilde {\cal {G}}} (r_{0}) \left[
(p+1) \frac {R''(r_{0})}{R(r_{0})}
\right. \nonumber \\ & & \left. \left.
- (n+p)
\left( \frac {R'(r_{0})}{R(r_{0})} \right) ^{2}
- \frac {f_{p+1}}{f_{p}} \frac {R'(r_{0})}{R(r_{0})}
\right]
\right\} ;
\label{eq:K}
\end{eqnarray}
and
\begin{eqnarray}
{\cal {G}}(r) & = & (n-1) Q(r) - T(r);
\nonumber \\
{\tilde {\cal {G}}} (r) & = & (n+1) Q(r) - T(r).
\label{eq:Gdef}
\end{eqnarray}

We therefore have that $T_{ {\cal {V}} {\cal {V}} }$ is finite at the horizon, while
\begin{eqnarray}
T_{ {\cal {W}} {\cal {W}} } & = &
\frac {1}{2} \left[ K_{1} (r-r_{0}) +O(r-r_{0})^{2} \right]
 f \left[ \psi ' (-{\cal {W}}) \right] ^{2}
 \nonumber \\  & &
 -\frac {\left[ \psi '( - {\cal {W}} )\right] ^{2} }{2 R^{(n-1)}}
\left[ Z+X \right] .
\label{eq:twwhor}
\end{eqnarray}
The second term in (\ref{eq:twwhor}) is
$O(r-r_{0})^{-2}$ as $r\rightarrow r_{0}$ for a non-extremal horizon ($p=1$),
$O(r-r_{0})^{-4}$  for a doubly coincident horizon ($p=2$)
and $O(r-r_{0})^{-6}$ for the ``ultra-cold'' black hole with $p=3$.
Therefore, $T_{ {\cal {W}} {\cal {W}} }$ will diverge severely unless $X+Z=0$.

Even if this is the case, $T_{ {\cal {W}} {\cal {W}} }$ will still be divergent
if $p>1$ as the first term in (\ref{eq:twwhor}) is $O(r-r_{0})^{-p+1}$.
The only way for the RSET to be regular on an extremal horizon is if, as well as imposing $X+Z=0$, we also have
$K_{1}=0$ (for $p=2$) and, in addition $K_{2}=0$ if $p=3$.
It is clear that these are strong constraints on the RSET, and, in general, it is unlikely
that either $K_{1}$ or $K_{2}$ will vanish, so the RSET will be divergent at an
extremal horizon.
We will show in section \ref{sec:2dext} that, in the simpler two-dimensional case, when $n=1$,
these conditions cannot be satisfied at an extremal horizon.
It is likely that this result extends to higher dimensions, but a full computation of the unknown
function $Q$ is required in this case.

The analysis proceeds similarly for a horizon where ${\cal {W}}$
is finite and non-zero, but ${\cal {V}}$ vanishes.
In this case, we require $X-Z=0$ in order for $T_{ {\cal {V}} {\cal {V}} }$ to be regular,
and, in addition, $K_{1}=0$ for an extremal horizon, with $K_{2}=0$ as well if the horizon is
ultra-extremal.

Most of the extremal black hole geometries shown in \cite{cardoso}
have both past and future extremal horizons, and, in order for the RSET to be finite on
both, it must be the case that $X=0=Z$ and $K_{1}=0$ (with $K_{2}$ also zero if $k=3$).
However, the ``Nariai''-type black hole (see \cite{cardoso}) is
different in that it has a future extremal horizon but no past horizon.
In this case the criteria are less stringent, and the RSET will be regular if $X+Z=0$
and $K_{1}=0$ (note that the horizon in this case is not ultra-extremal).

Typically, horizons where ${\cal {V}}=0$ will correspond to past horizons and
those with ${\cal {W}}=0$ will correspond to future horizons.
Assuming this to be the case,
we summarize our results from this section in table \ref{tab:nonext}, which gives the conditions on
the constants $X$ and $Z$ for the RSET to be regular on the possible combinations
of future (${\cal {H}}^{+}$) and past (${\cal {H}}^{-}$) non-extremal horizons.
These conditions are necessary and sufficient for regularity on non-extremal horizons;
for extremal horizons they are necessary but not sufficient as we also have the additional
requirements for the $K_{i}$ to vanish as outlined above.
\begin{table}
\begin{tabular}{|c|c|c|}
\hline
  Conditions on & \multicolumn{2}{c|}{Regular on}  \\
$X$ and $Z$ & ${\cal {H}}^{-}$
 & ${\cal {H}}^{+}$
     \\
\hline
  $X=Z=0$
& Yes & Yes
\\ \hline
 $X=Z\neq 0$ & Yes & No \\
\hline
$X=-Z\neq 0$ & No & Yes \\
\hline
 $ X\neq\pm Z$ & No & No \\
\hline
 \end{tabular}
 \caption{Conditions for the RSET to be regular on a non-extremal
 black hole event horizon.}
 \label{tab:nonext}
\end{table}
Table \ref{tab:nonext} gives us exactly the same results as in \cite{cf}: the only way that the RSET can be
regular on both the future and past event horizon is if $X=Z=0$, so that we have a time-symmetric
state which represents the Hartle-Hawking vacuum \cite{hh}.
For the Boulware vacuum \cite{b} we have $X\neq \pm Z$ and the RSET is divergent on both
event horizons.
In this case the RSET should be time-reversal symmetric so we set $X=0$ but $Z\neq 0$.
Finally, for the Unruh vacuum \cite{u}, the RSET is regular on the future but not the
past event horizon, so that $X=-Z\neq 0$, and we have a non-zero $P$ (\ref{eq:PA}), which
represents the outgoing Hawking flux.
The last case, in which the RSET is regular on the past but not the future event horizon,
may represent a ``future'' Unruh state (see, for example, \cite{aco}) in which there is an
ingoing flux of radiation but none emitted.
We do not consider this possibility further.

\section{Distinct event and cosmological horizons}
\label{sec:eventandcosmo}

We now turn to the case in which the geometry has two distinct, non-extremal, horizons
(the prototype being the Reissner-Nordstr\"om-de Sitter black hole, with an event and cosmological
horizon at $r_{+}$ and $r_{++}$ respectively).
We consider the region between the horizons
and use two sets of Kruskal co-ordinates, $V_{+}$, $W_{+}$ and $V_{++}$, $W_{++}$,
as discussed in section \ref{sec:kruskal}.
Near the event horizon, we have the same results as in the previous subsection,
for the co-ordinate system $V_{+}$, $W_{+}$, leading to table \ref{tab:nonext}.

Fixing $r_{a}=r_{+}$ as above, we find, near the cosmological horizon,
\begin{equation}
J(r_{+},r) = J(r_{+},r_{++}) + O(r-r_{++})^{2} ,
\end{equation}
which gives, near the future cosmological horizon ${\cal {C}}^{+}$,
\begin{eqnarray}
T_{V_{++}}^{W^{++}} & = &
\left[ {\tilde {Z}}+X \right]
\left[ C_{0} (r-r_{++})^{-2} + C_{1} \left( r-r_{++} \right) ^{-1} \right]
\nonumber \\ & &
+O(r-r_{++})^{0} ;
\nonumber \\
T_{W_{++}}^{V_{++}} & = &
\left[ {\tilde {Z}}-X\right]
\left[ D_{0} + D_{1} \left( r-r_{++} \right)  \right]
+O(r-r_{++})^{2} ;
\label{eq:cosmocond1}
\end{eqnarray}
where
\begin{equation}
{\tilde {Z}} = Z+J(r_{+},r_{++}) ;
\end{equation}
whilst, near the past cosmological horizon ${\cal {C}}^{-}$:
\begin{eqnarray}
T_{V_{++}}^{W_{++}} & = &
\left[{\tilde {Z}}+X \right]
\left[ {\tilde {C}}_{0} + {\tilde {C}}_{1}
\left( r-r_{++} \right)  \right]
+O(r-r_{++})^{2} ;
\nonumber \\
T_{W_{++}}^{V_{++}} & = &
\left[ {\tilde {Z}}-X \right] \left[ {\tilde {D}}_{0} (r-r_{++})^{-2}
+ {\tilde {D}}_{1} \left( r-r_{++} \right) ^{-1} \right]
\nonumber \\ & &
+O(r-r_{++})^{0} ;
\label{eq:cosmocond2}
\end{eqnarray}
where the $C$, $D$, ${\tilde {C}}$ and ${\tilde {D}}$ are fixed, non-zero constants.
All other components of the RSET are automatically regular across the cosmological horizon.
Using (\ref{eq:cosmocond1},\ref{eq:cosmocond2}), we can build up a table of conditions
similar to table \ref{tab:nonext}, but involving ${\tilde {Z}}$ rather than $Z$.
Combining all the possible behaviours at the event and cosmological horizons, one could
build up a long list of different combinations, although most of these will not be
physically relevant.
Instead, we list in table \ref{tab:cosmo} the properties of the physically relevant
static states (which are all time-reversal symmetric).
\begin{table}
\begin{tabular}{|c|c|c|c|}
\hline
 State & Conditions on & \multicolumn{2}{c|}{Regular on}  \\
& $X$, $Z$ and ${\tilde {Z}}$ & ${\cal {H}}$
 & ${\cal {C}}$      \\
\hline
Boulware & $ X=0\neq Z$, ${\tilde {Z}}\neq 0$ & No  & No  \\
\hline
Hartle-Hawking & $X=Z=0$, ${\tilde {Z}}\neq 0$ & Yes  & No  \\
\hline
Gibbons-Hawking & $X={\tilde {Z}}=0$, $Z\neq 0$ & No & Yes  \\
\hline
Regular &  $X=Z={\tilde {Z}}=0$ & Yes  & Yes  \\
\hline
 \end{tabular}
 \caption{Properties of the physically relevant static states on
 event and cosmological horizons.}
 \label{tab:cosmo}
\end{table}

Each of the two horizons, the event and cosmological horizon, will have an intrinsic
temperature associated to it, given by $\kappa /2\pi $, where $\kappa $ is the surface
gravity of that horizon (\ref{eq:kappa}).
If we therefore define a zero-temperature, ``Boulware'' state, then we would expect
that it will be divergent on both the event and cosmological horizons.
This state will be time-reversal symmetric, so $X=0$, but $Z$ and ${\tilde {Z}}$ are
unrestricted.

Secondly, we may consider a ``Hartle-Hawking'' state, which is a finite temperature state
at the same temperature as the event horizon, but, in general, will not be at the same
temperature as the cosmological horizon.
Therefore, in analogy with the Hartle-Hawking state for an asymptotically flat black hole,
we expect that this state will be regular on the event horizon.
However, due to the temperature difference between the event and cosmological horizons,
we expect this state to be divergent on the cosmological horizon.
As in the previous section, this means that $X=0=Z$ but leaves ${\tilde {Z}}$ unrestricted.

In a similar way, a ``Gibbons-Hawking'' \cite{gibbons} state, at the same temperature as
the cosmological horizon, will be regular on the cosmological horizon but divergent
on the event horizon in general.
Therefore, in this case, $X=0={\tilde {Z}}$ but $Z$ is arbitrary.

Although, in general, the temperatures of the event and cosmological horizons will be
different, there are special cases in which they are the same.
A good example of this is the ``lukewarm'' Reissner-Nordstr\"om-de Sitter
black hole \cite{romans,mellor}.
In this case, it is reasonable to suppose that a state at that temperature might
be regular on both the event and cosmological horizons.
This places great restrictions on the stress-tensor: it means that $X=0=Z$, and,
in addition ${\tilde {Z}}=0$, or, equivalently, $J(r_{+},r_{++})=0$, that is,
\begin{eqnarray}
0 & = & \frac{1}{2fR^{(n-1)}}
\int_{r_{+}}^{r_{++}}
\left[(n-1)Q-T \right]
f'R^{(n-1)} dr
\nonumber \\ & &
-\frac{(n-1)}{fR^{(n-1)}}
\int_{r_{+}}^{r_{++}}
QfR^{(n-2)} R' dr .
\label{eq:Qcond}
\end{eqnarray}
We will show in section \ref{sec:2d} that this condition is in fact satisfied for
the ``lukewarm'' black hole in two dimensions, when the unknown function $Q(r)$ is absent.
For four or higher dimensions, it is a non-trivial question as to whether a state can be
constructed for a ``lukewarm'' black hole such that (\ref{eq:Qcond}) holds, to which
we plan to return in the future \cite{phi2ren}.
On the other hand, the Kay/Wald theorem \cite{kay} tells us that there is no static state on
the Schwarzschild-de Sitter space-time which is regular on both the event and cosmological horizons.
In this case, equation (\ref{eq:Qcond}) does not hold.

We have not mentioned in this section the equivalent of the ``Unruh'' state for
this type of black hole space-time.
The reason for this is that the
state which represents the emission of Hawking radiation by a black hole
formed by gravitational collapse in de Sitter space is not static \cite{sds,choudhury}.

We comment that, although our analysis in this section has focussed on the case
of an event and cosmological horizon, similar considerations apply to
any two distinct horizons, such as an event and inner horizon.
In each case, a stringent integral condition similar to (\ref{eq:Qcond})
must be satisfied if the RSET is to be regular on both horizons.
A similar condition has been found for the regularity of the RSET on the inner horizon
of a Kerr black hole \cite{hiscockkerr}.

\section{Two-dimensional examples}
\label{sec:2d}

In this paper we are primarily concerned with higher-dimensional black holes,
but it is informative to study two-dimensional examples as in this case there is
no unknown function $Q$.
We consider two particular cases: (i) distinct, non-extremal event and cosmological horizons, and
(ii) extremal horizons.

\subsection{Distinct event and cosmological horizons}
\label{sec:2dEC}

In this subsection we consider a two-dimensional black hole with two distinct horizons, corresponding
to an event horizon and a cosmological horizon.
We will show that the RSET can only be regular on both horizons if the temperatures of the two horizons are equal.

From section \ref{sec:eventandcosmo}, the RSET can only be regular on both the event and cosmological
horizons if equation (\ref{eq:Qcond}) holds.
In two dimensions, this reduces to
\begin{equation}
0  = -  \frac{1}{2f}
\int_{r_{+}}^{r_{++}}
f' T  \, dr .
\label{eq:2dcond}
\end{equation}
In two dimensions, the trace anomaly $T$ is given by \cite{christensen}
\begin{equation}
T = \alpha {\cal {R}},
\label{eq:2dT}
\end{equation}
where ${\cal {R}}$ is the Ricci scalar of the two-dimensional metric (\ref{eq:metric})
and $\alpha $ is a constant, independent of the space-time geometry but dependent on the
spin of the quantum field under consideration.
For a conformally coupled quantum scalar field, for example,
\begin{equation}
\alpha = \frac {1}{24\pi }.
\end{equation}
In the two-dimensional case, the Ricci scalar is
\begin{equation}
{\cal {R}} = -f''(r).
\end{equation}
The integral (\ref{eq:2dcond}) then becomes
\begin{eqnarray}
0 & = &
\frac {\alpha }{2f} \int _{r_{+}}^{r_{++}} f' f'' \, dr
\nonumber \\
 =
\frac {\alpha }{4f} \left[ f'(r_{++})^{2} - f'(r_{+})^{2} \right]
\nonumber \\
 =
\frac {\alpha }{f} \left[ \kappa _{++}^{2} - \kappa _{+}^{2} \right] ;
\end{eqnarray}
using (\ref{eq:kappa}), where $\kappa _{++}$ and $\kappa _{+}$ are the
surface gravities of the cosmological and event horizons, respectively.
Therefore it must be the case that $\kappa _{++}=\kappa _{+}$ if the
RSET is regular on both the horizons.
Since the temperatures of the horizons are proportional to the surface gravities,
therefore the event and cosmological horizons have the
same temperature.
This occurs for ``lukewarm'' Reissner-Nordstr\"om-de Sitter black holes \cite{romans,mellor}.
Therefore, for two-dimensional ``lukewarm'' black holes, the RSET will be regular on both
the event and cosmological horizons if $X=0=Z$.
To extend this result to higher dimensions requires a computation of the function $Q(r)$, to
which we will return in the future \cite{phi2ren}.

\subsection{Extremal horizons}
\label{sec:2dext}

For extremal horizons, if the RSET is to be regular across the event horizon, as well
as the conditions on the constants $X$ and $Z$ outlined in table \ref{tab:nonext},
we also have conditions on the constants $K_{i}$ given in equation (\ref{eq:K}).
In two dimensions, the quantity ${\cal {G}}$ in (\ref{eq:Gdef}) reduces to
\begin{equation}
{\cal {G}} = -T
=\alpha f'',
\end{equation}
and therefore the constants $K_{i}$ become
\begin{eqnarray}
K_{1} & = &
-\frac {\alpha }{2(p+1)} f'''(r_{0});
\nonumber \\
K_{2} & = &
- \frac {\alpha }{2(p+1)(p+2)} \left[
 (p+1) f''''(r_{0})
\right. \nonumber \\ & & \left.
- \frac {f_{p+1}}{f_{p}} f'''(r_{0})
\right] .
\end{eqnarray}
For an extremal horizon, we have $p=2$, and the condition for regularity is that $K_{1}=0$.
However, for an extremal horizon, by definition $f'''(r_{0})\neq 0$ and so $K_{1} \neq 0$.
Therefore the RSET for a static state must diverge on an extremal horizon.
On the other hand, for an ultra-extremal horizon, we have $p=3$ and by definition $K_{1}=0$.
In this case, however, we also have the additional condition that $K_{2}$ must vanish if the
RSET is to be regular.
In the ultra-extremal case we have
\begin{equation}
K_{2} = -\frac {\alpha }{10} f''''(r_{0}) \neq 0,
\end{equation}
and once again it must be the case that the RSET diverges for a static state.

The question of the regularity of the RSET on extremal horizons has been particularly controversial in the literature
(see, for example, \cite{extremal}).
Our results here are in agreement with the general current consensus in the literature \cite{extremal},
namely that, if the RSET is to be
regular across an extremal event horizon, the state it describes must be non-static.

\section{Conclusions}
\label{sec:conc}

In this paper, we have examined the RSET for static quantum states on $(n+1)$-dimensional,
static, black hole space-times.
We have generalized the analysis of \cite{cf}, solving the conservation equations in this case.
The RSET is then given in terms of one unknown function of the radial co-ordinate $r$, and two unknown
constants $X$ and $Z$.
We have used generalized Kruskal-like co-ordinates, following \cite{liberati}, to study the behaviour of the RSET
near a horizon, and have derived conditions for the RSET to be regular there.
We hope that our results will be of use for full computations of the RSET on either higher-dimensional black holes,
or on black holes with a complicated horizon structure.
In particular, for a wide class of black hole space-times, and Hartle-Hawking-like quantum states,
the RSET is given by a single unknown function of the
radial co-ordinate $r$, which reduces the amount of computation required.

\begin{acknowledgements}
We would like to thank Adrian Ottewill for numerous invaluable discussions.
EW would like to thank the School of Mathematics and Statistics, University of
Newcastle-upon-Tyne for hospitality while part of this work was undertaken.
The work of EW is supported by UK PPARC, grant reference number PPA/G/S/2003/00082,
while that of PMY is supported by EPSRC UK.
\end{acknowledgements}

\end{document}